\documentclass[aps,twocolumn,showpacs]{revtex4}
\usepackage{amssymb}
\usepackage{graphicx}
\usepackage{amsmath}

\begin{document}

\title{Hardcore bosons on checkerboard lattices near half filling:
geometric frustration, vanishing charge order and fractional phase}
\date{\today}
\author{Yi Zhou}
\affiliation{Max-Planck-Institut f\"ur Physik komplexer Systeme \\
N\"othnitzer Str.38, 01187 Dresden, Germany}
\altaffiliation[Present address: ]{Department of Physics, Hong Kong University of Sicence and Technology, Clear
Water Bay, Kowloon, HongKong}

\begin{abstract}
We study a spinless hardcore boson model on checkerboard lattices by Green
function Monte Carlo method. At half filling, the ground state energy is
obtained up to $28\times 28$ lattice and extrapolated to infinite size, the
staggered pseudospin magnetization is found to vanish in the thermodynamic
limit. Thus the $(\pi,\pi)$ charge order is absent in this system. Away from
half filling, two defects induced by each hole (particle) may carry
fractional charge ($\pm e/2$). For one hole case, we study how the
defect-defect correlation changes with $t/J$, which is the ratio between the
hopping integral and cyclic exchange, equals to $V/2t$ when $V\gg t$.
Moreover, we argue that these fractional defects may propagate independently
when the concentration of holes (or defects) is large enough.
\end{abstract}

\maketitle

\section{Introduction}

During the past decade materials experiencing geometric frustration has been
a topic of much interest, both on experimental and theoretical sides \cite
{Frustration}. The rich behavior in such systems is due to a large
ground-state degeneracy, which renders them highly unstable to
perturbations. A well known 3D frustrated structure is the pyrochlore
lattice which is the backbone structure shared by many physically realistic
materials. One may think of checkerboard lattices as a 2D analog of
pyrochlore lattices. This is the reason why a checkerboard lattice is often
considered in theoretical studies instead of the more realistic pyrochlore
structure.

\begin{figure}[tbph]
\includegraphics[width=1.6in]{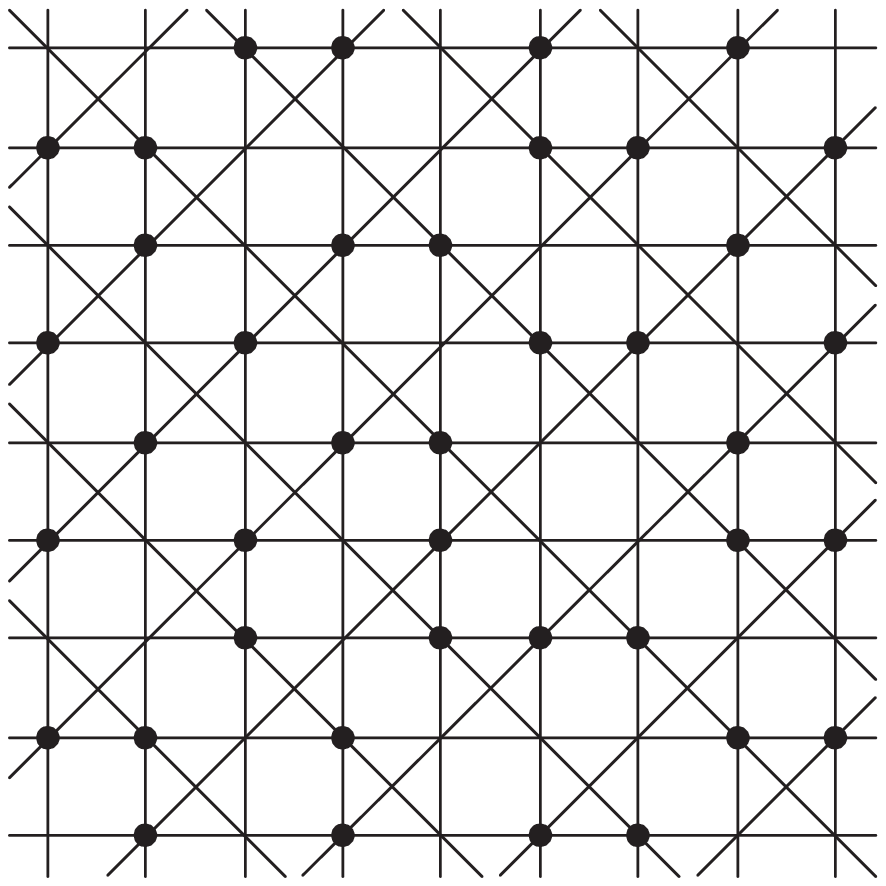} %
\includegraphics[width=1.6in]{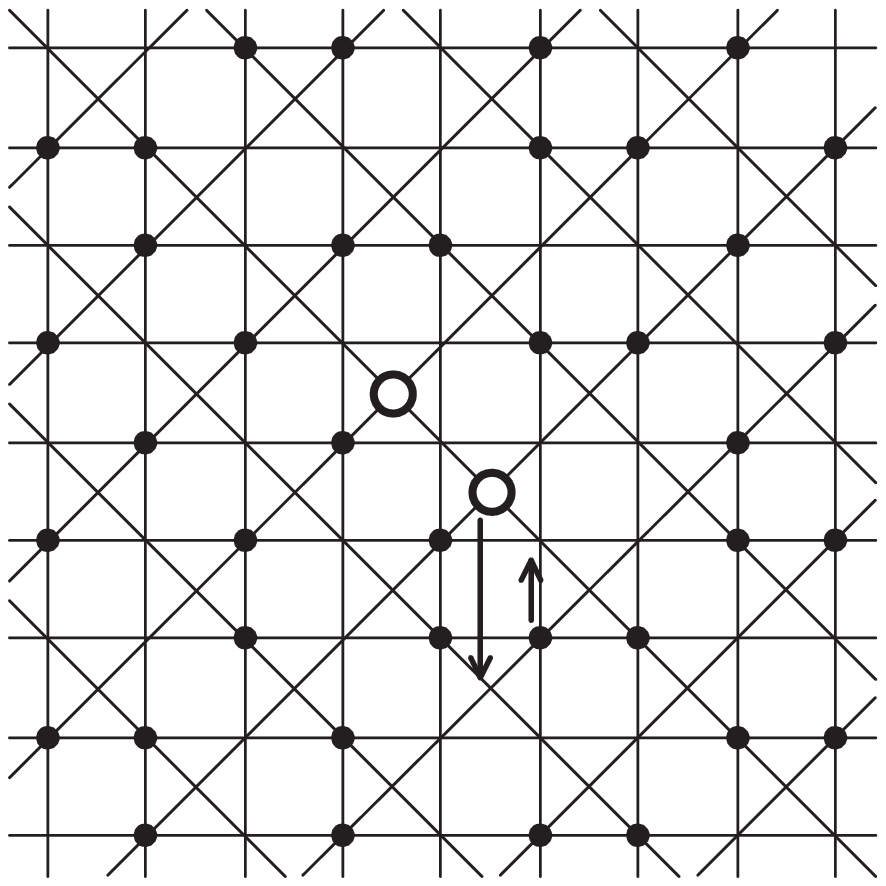}
\caption{Left: A typical configuration subject to the tetrahedron rule at
half filling. Right: Two defects induced by one hole may be considered as
point like fractional excitations, whose motion is driven by hopping
Hamiltonian.}
\label{fig.cfg}
\end{figure}

Up to now most of work has been devoted to an understanding of the magnetic
properties of such frustrated lattices, while the charge degrees of freedom
is waiting for more attention. Actually, the charge degrees of freedom may
exhibit fascinating physical properties in frustrated systems such as
pyrochlore and checkerboard lattices \cite{Fulde2002,Runge2004}. Recently,
one of the fascinating predictions on charge degrees of freedom has been
proposed, say, the geometric frustration present on the pyrochlore lattices
may give rise to fractional charges in two or three dimensions based on
strong nearest neighbor repulsion close to half-integer filling. This
proposal comes from the so called "tetrahedron rule" which is firstly stated
explicitly by Anderson \cite{Anderson1956} to explain the metal-insulator
transition in the spinel Fe$_{3}$O$_{4}$ \cite{Verwey1941}, where the
observed entropy reduction is much less than expected from electrons without
the strong short range correlation. Both pyrochlore and checkerboard
lattices are made up of corner sharing units which are tetrahedron or
crisscrossed plaquettes. The nearest neighbor repulsion will be minimized
when each of the corner sharing units contains only two particles. If the
kinetic energy can be neglected, the system will possess a large
ground-state degeneracy, $(4/3)^{(3N/4)}$ for checkerboard lattices \cite%
{Lieb1967}. Taking (putting) one particle from (into) the system, two
tetrahedra (plaquettes) will emerge, each of them contains an extra particle
(or hole). We will call this kind of tetrahedra (plaquettes) "defects". A
demonstration of the configuration subject to "tetrahedron rule" and the
formation of "defects" is shown in Fig.\ref{fig.cfg}. Provided that the
perturbation such as kinetic energy can be neglected, these two defects
induced by one particle (hole) will propagate independently. If a particle
carries charge $e $, one of the defects will carry charge $e/2$. Thus, such
defects are "fractional". Note that the above argument is valid for both
hardcore bosons and fermions as well as spinless and spin-$S$ system. On the
other hand, current laser cooling and cold atom technique make it possible
to realize these systems on some artificial optical lattices \cite%
{Zoller2005}.

However, the virtual processes induced by kinetic energy will lift the high
degeneracy and lead to different ground states. Because of high degeneracy,
any small perturbation may change the low energy states violently. To study
this kind of quantum effects in spinless fermionic system on checkerboard
lattices, Ref. \cite{Runge2004} used exact diagonalization (ED) technique up
to 32 sites. As a related problem to hardcore bosons, XXZ Heisenberg model
in the Ising limit was studied by ED on small size lattices too \cite%
{Shannon2004}. The ground state was identified as a nonmagnetic state of
resonating square plaquettes. Due to the limit of small size, there is lack
of direct evidence to address the issue of confinement or deconfinement.

In this paper, we will study spinless hardcore bosons by Green function
Monte Carlo (GFMC) method which can give precise results on larger lattices
(up to $28\times 28$). The ground state energy is obtained with high
accuracy and defect-defect correlation is present, which makes sense only on
larger lattices. Also, we will inspect the issue of charge order. The
outline of the paper follows. In section II, we introduce the extended
Hubbard model and derive its effective Hamiltonian subjected to tetrahedron
rule in the strong repulsion limit. The relations between this model and XXZ
Heisenberg model is discussed. In section III, a brief discussion of GFMC method
is present, we also compare some results on small lattices with exact
diagonalization. Section IV contains the main numerical results. The final
section is devoted to conclusions.

\section{The effective Hamiltonian}

We consider strong on-site repulsion $U$ and nearest-neighbor repulsion $V$
between spinless hardcore bosons on checkerboard lattices, where the
intra-site interaction has been assumed to be a higher energy scale and
ignored. Then the extended Hubbard Hamiltonian is of the form 
\begin{equation}
H=-t\sum_{\langle ij\rangle }(b_{i}^{\dagger }b_{j}+h.c.)+V\sum_{\langle
ij\rangle }n_{i}n_{j}+U\sum_{i}n_{i}^{2},  \label{HtV}
\end{equation}%
where $b_{i}(b_{i}^{\dag })$ denote annihilation (creation) operators at
site $i$, $n_{i}=b_{i}^{\dag }b_{i}$, and $\langle ij\rangle $ refers to a
pair of nearest neighbor. We shall assume that $U\gg V,t$ is large enough to
ensure the non-double occupancy condition, only empty and singly occupied
sites ar considered. Then we will focus on the strong interaction regime $%
V\gg t>0$ around half filling where the average occupation number per site
is $\left\langle n_{i}\right\rangle =1/2$, more precisely, half of the sites
are occupied and the other half of the sites are empty. In that case the
tetrahedron rule is imposed by the strong nearest neighbor repulsion, 
\begin{equation}
\sum_{i\in \boxtimes }n_{i}=2,  \label{t-rule}
\end{equation}%
where $i$ belongs to a same crisscrossed plaquette. In this way, the
inter-site repulsion will be minimized. Thus, in the limit $V\gg t$, we
obtain the effective low energy Hamiltonian in the subspace restricted by
the condition (\ref{t-rule}), 
\begin{equation}
H_{J}=-J\sum_{\square }b_{i_{1}}^{\dagger }b_{i_{2}}b_{i_{3}}^{\dagger
}b_{i_{4}}+h.c.,  \label{HJ}
\end{equation}%
where $J=2t^{2}/V$, and $\square $ denotes a four-site loop without
crisscross, formed by sites $i_{1}i_{2}i_{3}i_{4}$. Away from half filling,
we consider the hole doping only, due to the particle-hole symmetry at
half-filling. Then the tetrahedron rule has to be modified as the following, 
\begin{equation}
\sum_{i\in \boxtimes }n_{i}\leq 2,  \label{mt-rule}
\end{equation}%
and the effective Hamiltonian in this subspace is given by 
\begin{equation}
H_{eff}=-t\sum_{\langle ij\rangle }(b_{i}^{\dagger }b_{j}+h.c.)+H_{J}.
\label{Heff}
\end{equation}%
It means that if we take one particle from the half filling system, there
will emerge two defects where $\sum_{i\in \boxtimes}n_{i}=1$ and $\sum_{i\in
\boxtimes}n_{i}=2$ elsewhere. The hopping terms (proportional to $t$) of (%
\ref{Heff}) will change the defects position but $H_{J}$ will not. Although $%
t/J=V/2t\gg 1/2$, we can generally consider an effective model in which the
ratio $t/J$ ranges from zero to positive infinity. As pointed out in the
former references \cite{Fulde2002,Runge2004,Shannon2004}, if the virtual
process at order $J$ can be neglected, these two defects will propagate as
independent fractional objects. However, quantum effects such as the cyclic
exchange of the order of $J$ may or may not confine these fractional defects.

A hardcore boson model can be mapped onto a spin-$1/2$ model in general. The
occupation number $n_{i}$ corresponds to the spin component $S_{i}^{z}+\frac{
1}{2}$, and creation (annihilation) operators $b_{i}^{\dag }\left(
b_{i}\right) $ correspond to $S_{i}^{+}\left( S_{i}^{-}\right) $. The
according spin-$1/2$ model of (\ref{HtV}) is an XXZ Heisenberg model. It is
noted that this XXZ Hamiltonian is \textit{different} from an usual XXZ
model by the \textit{minus} sign before spin flip terms, which can \textit{%
not} be mapped to a positive one by an unitary transformation on a \textit{%
frustrated} lattices. The related effective Hamiltonian in the Ising limit
is a cyclic exchange which was studied by Shannon \textit{et al}. \cite%
{Shannon2004} through ED on small lattices. The authors argued that spinons
(defects) in the above XXZ model will be confined due to the ground state
correlation \cite{confinement1}. In this paper we confirm this argument
through calculating the defect-defect correlation directly by GFMC method up
to $24\times 24$ lattices which are large enough to suppress the size
effect. If only one hole is present the two induced defects will be confined
together. However, as concentration of holes and the ratio $t/J=V/2t$
increase these defects may behave as independent point like objects.

\section{Ground state wave function and GFMC approach}

Note that both Hamiltonians (\ref{HJ}) and (\ref{Heff}) have only
non-positive off-diagonal matrix elements in the Fock representation $%
\left\vert n_{i1}n_{i2}\cdots \right\rangle $. It implies that the many-body
boson wave function can be chosen to be non-negative everywhere in the
ground state. This property of the wave function will
be crucial in applying the GFMC method to this system.

GFMC is a general scheme for finding the lowest eigenvalue of an operator. A
trial eigenvector is subjected to a time evolution whose effect is to
enhance those components of the solution with the lower eigenvalues of the
operator. The ground state energy can be obtained as a mixed estimate. 
We choose the trial state $\left\vert \psi _{T}\right\rangle $ as an equal amplitude
superposition in an invariant subspace of Hamiltonian which contains the
ground state. For a local operator $O$ whose matrix elements satisfy $\left\langle
R\right\vert O\left\vert R^{\prime }\right\rangle =\delta \left( R-R^{\prime
}\right) O\left( R\right) $, such as density-density correlation function,
which does not share eigenstates with the Hamiltonian, we use the
"forward-walking" technique, well known in many-body theory \cite%
{Binder1979,Chin1984,Kalos1966,Liu1974,Whitlock1979}, to measure the expectation
values. In this way, at the $n$th step of iteration, $O\left( R_{n}^{i}\right) $ is
evaluated for each configuration $R_{n}^{i}$ in the ensemble $\left\{
R_{n}\right\} $. For the remaining $m$ steps of the random walk, a record is
kept of the configuration of $\left\{ R_{n}^{i}\right\} $ from which each
subsequent ensemble member has evolved. At the end of the $m+n$th step, $%
\left\langle O\right\rangle $ is evaluated by forming the weighted summation 
\cite{Chin1984} 
\begin{equation*}
\left\langle O\right\rangle =\frac{\sum_{j}\psi _{T}\left(
R_{m+n}^{j}\right) O\left( R_{n}^{i\left( j\right) }\right) }{\sum_{j}\psi
_{T}\left( R_{m+n}^{j}\right) },
\end{equation*}%
where the notation $i\left( j\right) $ indicates that $i$ is the progenitor
of $j$. In this paper, we will calculate density-density correlation and
defect-defect correlation, both of them are local operators and this method
can be applied.

\subsection{Comparison with exact diagonalization}

GFMC is a very accurate method to obtain the ground state and low-lying
excited state property of many-body interacting systems. In the past it has
been applied successfully to the ground state properties of helium,
interacting electron gas, small molecules, Heisenberg model on 2D square
lattices, and lattice gauge theories \cite%
{Binder1979,Chin1984,Gross1989,Carlson1989,Trivedi1990}. To check the
precision of the GFMC method in our system, we will compare some results
from the ED with those from GFMC at small lattices with periodic boundary
condition (PBC). Hereafter we will use PBC in this paper.

The ground state energy at half filling and at one hole doped lattice are
given in Tables \ref{tab1} and \ref{tab2} respectively. The digits in the
bracket are statistics error bars.

\begin{table}[htbp]
\caption{The ground state energy $E/J$ at half filling.}
\label{tab1}
\begin{tabular}{|c|c|c|c|}
\hline
Size & 4$\times $4 & 4$\times $6 & 6$\times $6 \\ \hline
ED & $-4.47214$ & $-6.46995$ & $-9.47393$ \\ \hline
GFMC & $-4.4717(12)$ & $-6.4695(19)$ & $-9.4739(22)$ \\ \hline
\end{tabular}
\end{table}

\begin{table}[htbp]
\caption{The ground state energy $E/J$ of one hole doped $4\times 4$ lattice.}
\label{tab2}
\begin{tabular}{|c|c|c|}
\hline
$t/J$ & ED & GFMC \\ \hline
$0.02$ & $-3.52850$ & $-3.5290(9)$ \\ \hline
$0.2$ & $-4.91215$ & $-4.9119(16)$ \\ \hline
$2.0$ & $-22.92923$ & $-22.9310(16)$ \\ \hline
\end{tabular}
\end{table}

Density-density correlation $D(i,j)=D\left( i-j\right) \equiv \left\langle
n_{i}n_{j}\right\rangle -\frac{1}{4}$ at half filling on $6\times 6$ lattice
have been examined too. The results from ED and GFMC are present in Table
\ref{tab3} and \ref{tab4} respectively. Each of the two tables contains a matrix
$D(i_{x},i_{y})$ whose column and row indices $ i_{x},i_{y}=0,1,2,3$ correspond to 
the displacements along two directions respectively.

\begin{table}[htbp]
\caption{Density-density correlatoin $D(i_{x},i_{y})$ on $6\times 6$ lattice 
calculated by ED.}
\label{tab3}
\begin{tabular}{|c|c|c|c|}
\hline
0.25 & -0.13170 & 0.01933 & -0.02527 \\ \hline
-0.13170 & 0.07083 & -0.01590 & 0.02185 \\ \hline
0.01933 & -0.01590 & 0.01584 & -0.01920 \\ \hline
-0.02527 & 0.02185 & -0.01920 & 0.01998 \\ \hline
\end{tabular}
\end{table}

\begin{table}[htbp]
\caption{Density-density correlatoin $D(i_{x},i_{y})$ on $6\times 6$ lattice 
calculated by GFMC.}
\label{tab4}
\begin{tabular}{|c|c|c|c|}
\hline
0.25 & -0.1319(7) & 0.0196(12) & -0.0254(11) \\ \hline
-0.1319(7) & 0.0711(9) & -0.0161(10) & 0.0220(8) \\ \hline
0.0196(12) & -0.0161(10) & 0.0160(10) & -0.0193(7) \\ \hline
-0.0254(11) & 0.0220(8) & -0.0193(7) & 0.0200(7) \\ \hline
\end{tabular}
\end{table}



The staggered pseudospin magnetization (\ref{mag}) on a half filled $6\times
6$ lattice given by GFMC is $0.1038(8)$, while the result from ED is $%
0.10417 $.

From the above, one sees that GFMC algorithm is an effective method to deal
with spinless hardcore bosons on checkerboard lattices.

\section{Results}

Simulations were carried out for $L\times L$ lattices up to $L=28$ ($L=24$
for correlations). Time steps from 0.003 to 0.05 were used, depending on
lattice size and the ratio $t/J$, here we set $J=1$. In practice, we use
about $100L^{2}$ generations to reach the ground state distribution, then
iterate about $1000L^{2}$ generations to measure the physical quantities. To
avoid self-correlation and improve the efficiency, we make an expectation
value measurement only after every $L^{2}$ iterations. It is not
advantageous to perform measurements at still larger intervals, since the
measured generations are already nearly statistically independent. For
ergodicity we should control the population of random walkers large enough.
Otherwise they will be trapped in a higher energy state instead of the
ground state. According to our experience, keeping $L^{3}$ random walkers in
each generation is enough to ensure ergodicity on $L\times L$ lattice.

\subsection{Half filling: ground state energy and staggered pseudospin
magnetization}

\begin{figure}[tbph]
\includegraphics[width=3.8in]
{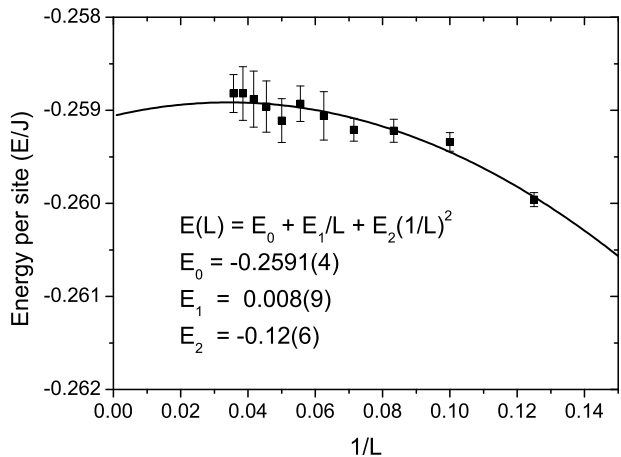}
\caption{Ground state energy per site at half filling and extrapolation to
thermodynamic limit. Values are shown for $L \times L$ lattices, $%
L=8,10,\cdots,28$.}
\label{E0}
\end{figure}

Firstly, we calculate the ground state energy at half filling which may
serve as a standard to compare with other analytical or numerical study in
this system. Fig. \ref{E0} shows ground state energies per site up to $%
28\times 28$ lattice. We extrapolated it to the thermodynamic limit through
the following formula, 
\begin{equation}
\frac{E(L)}{J}=E_{0}+\frac{E_{1}}{L}+\frac{E_{2}}{L^{2}}+O\left( \frac{1}{
L^{3}}\right),
\end{equation}
where $E_{0}=-0.2591(4)$, $E_{1}=0.008(9)$,$\,E_{2}=-0.12(6)$ and $L$ is the
linear size of lattices. 

\begin{figure}[tbph]
\includegraphics[width=3.8in]{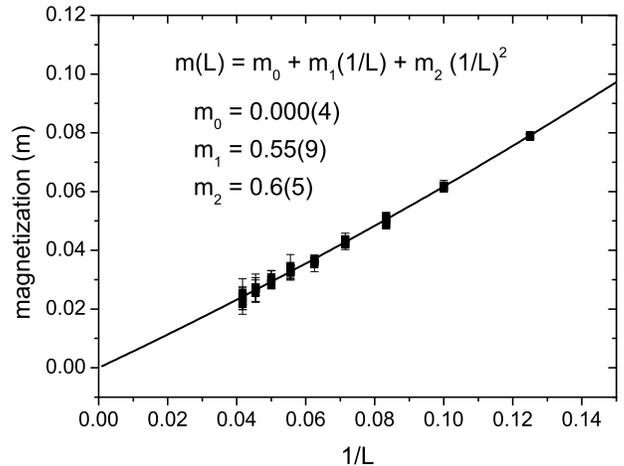}
\caption{Staggered pseudospin magnetization, Eq. (\protect\ref{mag}), on
finite size lattices at half filling and extrapolation to thermodynamic
limit. Values are shown for $L\times L$ lattices, $L=8,10,\cdots ,24$. Zero
result for $m_{0}$ means charge order is absent here.}
\label{mageps}
\end{figure}

Another issue that we shall address is whether there exists charge density
order in the thermodynamic limit. Although density-density correlation $%
D\left( i,j\right) $ exhibits $\left( \pi ,\pi \right) $ charge-modulation
patterns on small lattices with PBC, as it behaves in fermionic system \cite%
{Runge2004}, it may vanish in the thermodynamic limit. To answer this
question, we introduce the staggered pseudospin magnetization $m$, which is
defined as 
\begin{equation}
m=\frac{1}{2N}\sqrt{\left\langle \left( \sum_{i}\left( -1\right)
^{i}n_{i}\right) ^{2}\right\rangle },  \label{mag}
\end{equation}%
where $N$ is the number of sites on checkerboard lattices. The staggered
pseudospin magnetization has its name because if we map a hardcore boson
model to a spin-1/2 model this quantity is nothing but the staggered
magnetization. If the system is charge ordered, $m$ will not vanish in the
thermodynamic limit. By "forward walking" technique, we calculate the
pseudospin magnetization up to $24\times 24$ lattice and extrapolated it to
the thermodynamic limit by the following formula, 
\begin{equation}
m\left( L\right) =m_{0}+\frac{m_{1}}{L}+\frac{m_{2}}{L^{2}}+O\left( \frac{1}{%
L^{3}}\right) ,
\end{equation}%
with $m_{0}=0.000(4),$ $m_{1}=0.55(9)$ and $m_{2}=0.6(5)$. As shown in Fig.%
\ref{mageps}, it results in a vanishing charge order at $\left( \pi ,\pi
\right) $ in the thermodynamic limit. This result agrees with the claim of
non-magnetic phase in Ref. \cite{Shannon2004}. Similar conclusion was found
for spinless fermion \cite{Runge2004}. 

\subsection{One hole doping: ground state energy and defect-defect
correlation}

\begin{figure}[tbph]
\includegraphics[width=3.8in]{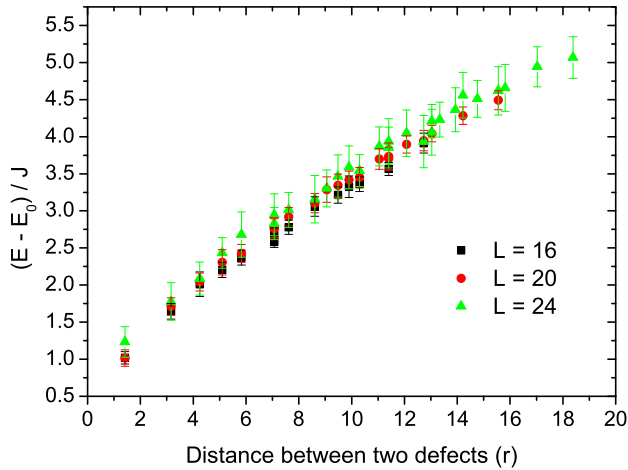}
\caption{(Color online) Cyclic exchange energy $H_{J}$ will increase
linearly when the distance between two defects increases in the case of $t=0$%
. Here $E_{0}$ is the ground state energy at half filling.}
\label{E0dop}
\end{figure}

One of the central questions is whether or not the defects will be confined
by the Hamiltonian (\ref{Heff}). To answer this question, we calculate the
ground state energy at one hole doping and $t=0$ at first. When $t=0$, the
two defects can not move away by hopping terms, so that we should calculate
the ground state energy with fixed defects. Fig. \ref{E0dop} shows the
numerical results for $L\times L$ lattice, $L=16,20,24$. It turns out that
the ground state energy will increase linearly as the distance between two
defects increases. The distance between two defects is defined as the
distance from one plaquette center to another in units of the lattice
constant $a$, say, it is $\sqrt{2}$ for two nearest neighbor plaquettes.

Even though we turn on the hopping term, the upper bound of the gained
kinetic energy is less than $2zt$, $z=6$ for checkerboard lattices is the
coordination number. Since the confinement potential increases linearly with
the distance between the two defects, the two defects will be confined with
an average distance $R$.

Now we turn to $t>0$ case and study how the defect-defect correlation will
change as the ratio $t/J$ varies. The defect-defect correlation $C\left(
p,q\right) =C\left( p-q\right) $ can be defined as: 
\begin{equation}
C\left( p,q\right) =\left\langle (2-\sum_{i\epsilon
p}n_{i})(2-\sum_{j\epsilon q}n_{j})\right\rangle ,  \label{cpq}
\end{equation}%
where $p$, $q$ denote the crisscrossed plaquettes. 
\begin{figure}[tbph]
\centering \includegraphics[width=3.8in]{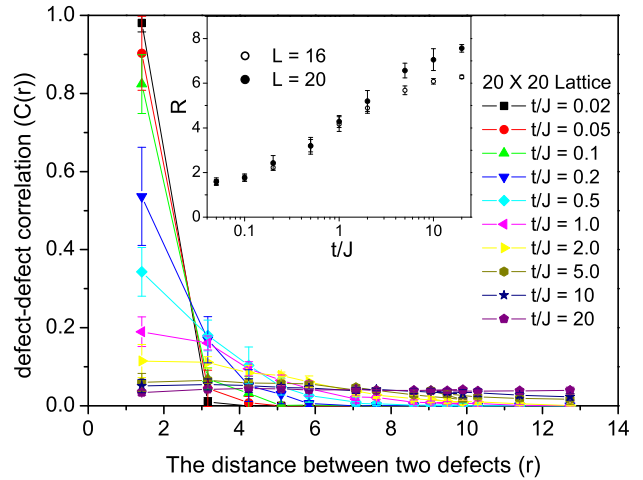} \label{cr}
\caption{(Color online) The defect-defect correlation on $20\times 20$
lattice. As the ratio $t/J$ increasing, it will become more and more
uniform. Inset: The average distance $R$ between the two defects
increases monotonically with $t/J$. The saturation is due to the size effect,
for a homogeneous distribution, the maximum $R$ will be 0.3826L.
Solid dots are plotted for $20\times 20$ lattice and circles for $16\times 16$ lattice.}
\label{cpqeps}
\end{figure}
The numerical result is present in Fig.\ref{cpqeps}. One notices that when $%
t/J$ is small, the two defects will be linearly bounded. As $t/J$
increasing, the defect-defect correlation will be more and more uniform in a
finite size lattice with periodic boundary condition. Hence we expect that
the average distance $R(t/J)$ between the two defects in the infinite
lattice will increase monotonically with $t/J$. However, as shown in the inset
of Fig.\ref{cpqeps}, it will reach the saturation on a finite lattice 
due to the size effect, the maximum will be $(\sqrt{2}+\ln (1+\sqrt{2}))L/6=0.3826L$ on
$L\times L$ lattice for a homogeneous distribution.

From the above we can conclude that the cyclic exchange $H_{J}$ provides a
linear confinement potential between two defects but the hopping term tends
to increase this average distance. If the concentration of defects $\delta
=2N_{hole}/N$ is large enough as to satisfy 
\begin{equation}
\delta \geq \frac{a^{2}}{\pi R^{2}(t/J)},
\end{equation}
when the average area occupied by one defect, $Na^2/(2N_{hole})$, is smaller
than the confinement area, $\pi R^2$, the defects will have a homogeneous
distribution on the infinity lattice instead of to be confined together in
couples. It implies that it is possible to treat these defects as
independent point like excitations.

\section{Conclusion}

In summary, we apply the GFMC algorithm to a spinless hardcore boson model
with strong nearest neighbor repulsion on checkerboard lattices near half
filling. To avoid any variational bias from the trial wave function, a
"forward walking" technique has been used to compute density-density
correlation and defect-defect correlations. At half filling, the ground
state energy is obtained and extrapolated to infinity size. It turns out
that the staggered pseudospin magnetization vanishes in the thermodynamic
limit, thus charge order at $\left( \pi ,\pi \right) $ is absent in this
system. Away from half filling, two defects induced by each hole (particle)
may carry fractional charge ($\pm e/2$). In the case of one hole doping, we
study how the defect-defect correlation changes with the parameter $t/J$,
which equals to $V/2t$ when $V\gg t$. The cyclic exchange is found to
provide a linear confinement potential between two defects, while the
hopping term as kinetic energy is tending to separate them away. Moreover,
we argue that these defects may propagate independently when the
concentration of holes (or defects) is large enough.

Finally, confinement (deconfinement) is a subtle issue, although we have
presented some evidences for the possibility of the existence of defects
which may carry fractional charge ($\pm e/2$) in this system, a lots of work
remain to be done. This issue should be treated by other analytical and
numerical methods. Especially, the effective field theories for the present
system are expected to describe the ground state and low lying excitations
well, thus a confident conclusion for confinement/deconfinement will be
achieved.

The author would like to thank Prof. P. Fulde for bringing his attention to
this field, also for his encouragement, stimulated discussion and critical
comment on the manuscript. Helpful discussions with F. Pollmann, E. Runge,
N. Shannon and Y. Zhang is acknowledged too.

\end{document}